\documentclass[showpacs,twocolumn,pre]{revtex4-1}

\usepackage{amsmath}
\usepackage{latexsym}
\usepackage{epsfig}
\usepackage{epstopdf}
\usepackage{makeidx}
\usepackage{amsfonts}

\begin{document}

\title{Ageing in Mortal Superdiffusive L\'{e}vy Walkers}
\author{Helena Stage}
\affiliation{School of Mathematics, The University of Manchester, Manchester M13 9PL, UK}

\begin{abstract}
A growing body of literature examines the effects of superdiffusive subballistic movement pre-measurement (ageing or time lag) on observations arising from single-particle tracking. A neglected aspect is the finite lifetime of these L\'{e}vy walkers, be they proteins, cells or larger structures. We examine the effects of ageing on the motility of mortal walkers, and discuss the means by which permanent stopping of walkers may be categorised as arising from `natural' death or experimental artefacts such as low photostability or radiation damage. This is done by comparison of the walkers' mean squared displacement (MSD) with the front velocity of propagation of a group of walkers, which is found to be invariant under time lags.
For any running time distribution of a mortal random walker, the MSD is tempered by the stopping rate $\theta$. This provides a physical interpretation for truncated heavy-tailed diffusion processes and serves as a tool by which to better classify the underlying running time distributions of random walkers. Tempering of aged MSDs raises the issue of misinterpreting superdiffusive motion which appears Brownian or subdiffusive over certain time scales.
\end{abstract}

\maketitle

\section{Introduction}
Superdiffusive motion has been experimentally observed in a wide range of scenarios \cite{superflow, superlight}, from human movement \cite{humans, supertravel} to bacterial swarming \cite{swarm} and intracellular transport \cite{supercell}, often by the tracking of individual walkers (be these proteins, humans, cells or otherwise) \cite{revspt, sptfix, single1, single2, single3}. Due to the complex nature of the transport, a common assumption made in the analysis of their motility is that the observed motion commences at the time of observation ($t=0$). 
That is, observation begins immediately after preparation of the system, such that the velocity $\vec{v}$ at which the walker is moving was sampled at the beginning of our measurements. If this is not the case, as occurs in a multitude of experimental setups \cite{agectrw, summaryreview}, the system is subject to \textit{time lag} or \textit{ageing}. 
The assumption of no ageing becomes especially problematic in cases where the walkers' movements are time-affected and have finite life spans and reproduction patterns, such that there is a finite time over which movement occurs.
\newline
In this work we consider the effects of such time lags on superdiffusive L\'{e}vy walkers with a finite lifetime in which they are motile, after which the walkers \textit{stop}. We further study the effects of replenishing these walkers, which either leads to a constant or increasing number of walkers. The physical interpretations of a walker stopping are manifold, but can be thought of as the `death' of the walker (with replenishing as `birth'). Death may be `natural' or induced by e.g. radiation damage during observation, as can be the case when utilising optical tweezers with light of high intensity \cite{tweezer2, tweezer1, tweezer3}. However, the end of a trajectory may also result from photobleaching \cite{bressloff, sedimentation, method1, method2, bleach2}, expiry of the walker fluorophore's ability to emit light as a result of photodissociation \cite{bressloff, sedimentation, single3, photostable}, or sedimentation aided by the probe size or mass \cite{sedimentation} of the walker in the observed sample. For \textit{in vivo} experiments, the walker may also bind to another compound in the system. `Mortal' thus simply refers to the finite duration of time over which the walker moves. The main question is the extent to which ageing and stopping affect the observed transport, which we shall gauge via the Mean Squared Displacement (MSD) and its time average (TAMSD). By producing theoretical predictions of the properties of these quantities, we obtain a test by which the empirically measured (TA)MSDs may be studied to ascertain whether such stopping effects are taking place.

Besides improving the accuracy of our models, the inclusion of time lags for mortal walkers leads to qualitatively different results. The implication of this is clear: for systems where time lag occurs, the nature of movement of each individual walker may be obfuscated if ageing and stopping of walkers are not considered. 
Over longer time scales one intuitively expects the effects of ageing to disappear but crucially, for a plethora of practical reasons, experiments may not be continuously conducted for long durations. Hence, if the time lag is longer than or comparable to the time scale of the experiment, it can significantly influence the observations. Similarly, if one is at liberty to discard any trajectory wherein the walker stops, the following analysis may seem superfluous. However, particularly for \textit{in vivo} experiments, the transport of a walker as it approaches another distinct region may be of interest, and naturally requires the previously ongoing process to stop. Examples include transport close to the cell membrane \cite{stop3}, the cell nucleus \cite{stop2} and neuronal transport \cite{stop1}. In the case of active motion, the time scales over which stopping occur can inform the loss of ATP or other energy sources.

\subsection*{What Is Age?}
\begin{figure}[h]
\includegraphics[scale=.5]{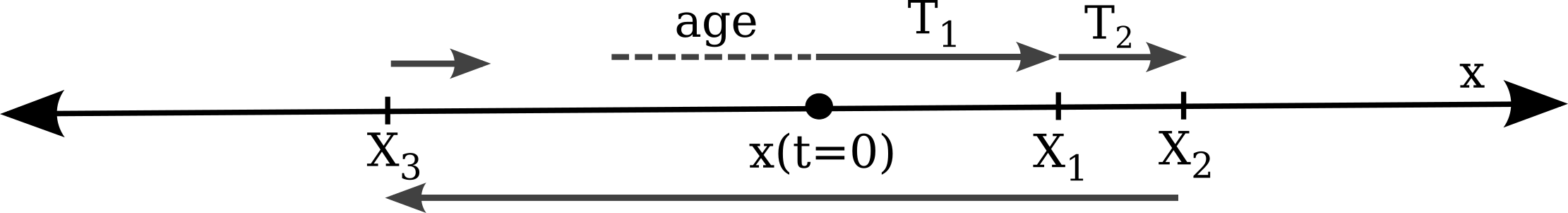}
\caption{An illustration of a random walk on the line. Observed movement happens with finite velocity $\pm v$ for some time $T_i$ (for the $i^{th}$ segment of the walk), starting from the initial position $x(t=0)$. The continuous movement leads to a time-dependent position $x(t)$. As shown with the full arrow, no age corresponds to the case where $t=0$ coincides with the time at which a running time is sampled for a new segment ($T_1$ sampled at $t=0$). Otherwise, for an aged process this run started some time before (as shown by the dotted line).
\label{fig: line-illus}}
\end{figure}

Let us now define what we mean by time lag or age in a random walk.
Consider a system of non-interacting random walkers moving on the real line with a given velocity $\vec{v}$ which takes values $\pm v$ moving to the right and left, respectively. The walkers move continuously in a certain direction for a time $T$, which we call the \textit{running time}, drawn from a probability density function (PDF) $\psi(\tau)$. Consequently, the probability of uninterrupted movement is given by the survival probability $\Psi(\tau)=\int_\tau^\infty\psi(u)du$ \cite{feller}. Similarly, the position of the walker is a random quantity $X(t)$ which for symmetric random walks has zero mean. So, a walker will draw a running time $T_1$ and move for a distance $\vec{X}_1=\vec{v}_1T_1$, after which a new running time and direction of movement will be sampled and a distance $\vec{X}_2=\vec{v}_2T_2+\vec{X}_1$ will be covered, etc.
For the PDFs we consider the mean running time $\left<T\right>$ as finite since the integral $\int_0^\infty\tau\psi(\tau)d\tau=\int_0^\infty\Psi(\tau)d\tau$ converges. 
For an illustration of this movement, see Figure \ref{fig: line-illus}.
If there is no ageing in the random walk, $T_1$ is sampled at $t=0$. However, if the system is lagged this movement will have commenced before observation at some time $t<0$ and thus the total running time of a run ending at $X_1$ has duration longer than the apparent value of $T_1$. Since each run is independent of the next, we need only consider the last run before observation in the case of time lagged movement.

The MSD $\left<X^2(t)\right>$ of this random walk considers the PDF $P(\vec{x},t)$ of a walker being at position $\vec{x}$ at time $t$ and is defined (for simplicity here in one dimension) as 
\begin{equation}
\left<X^2(t)\right>=\int_0^\infty x^2P(x,t)dx.
\end{equation}
Recall that for classical diffusion (as originally described by Fick's first Law) a diffusion coefficient $D$ leads to an MSD $\left<X^2(t)\right>=2Dt$ which grows linearly in time \cite{klafter, walkersrev, genctrw4}.

One can introduce a mean structural probability density $n(\vec{x},t,\tau)$ which describes the probability density of walkers found at position $\vec{x}$ at time $t$ with a given running time. That is, $n(\vec{x}, t,\tau)\Delta\tau$ gives the probability of walkers with running times in the interval $(\tau, \tau+\Delta\tau)$ assuming the population of walkers is conserved \cite{coxmiller, vlad02, SergeiBook}. Standard initial conditions which assume the current run started at observation $t=0$ are given by
\begin{equation}
n(\vec{x},0,\tau)=p_0(\vec{x})\delta(\tau).
\end{equation}
Under this assumption, all walkers initially have zero running time \cite{klafter, walkersrev}. However, if we are to consider systems which require preparation before observations can commence, movement will have begun before $t=0$ and will thus have non-zero running times. The introduction of our structural probability $n(\vec{x},t,\tau)$ makes the description of the random walk semi-Markovian as each run is independent of previous ones. For ageing systems we thus need only consider the last run commenced before $t=0$ (and not ending at that time).
We consider the following two types of time lag which may occur:
\begin{equation}
n(\vec{x},0,\tau)=\begin{cases}
p_0(\vec{x})\delta(\tau-\tau_0) &\tau_0\text{ initial running time,}\\
p_0(\vec{x})\frac{\Psi(\tau)}{\left<T\right>} &\text{Equilibrium state.}
\end{cases}
\label{eq: time-conditions}
\end{equation}
The first case describes a system wherein at $t=0$ all walkers have running times $\tau_0$, such that  movement began at $t\leq-\tau_0$ and has continued uninterrupted since $t=-\tau_0$. This is subtly different from a random walk wherein movement started at a time $t=-\tau_0$, but the walker may have changed direction in the interim period.
The second case of \eqref{eq: time-conditions} describes the walkers having reached an equilibrium state before measurements began \cite{coxmiller}. That is, a sufficiently long time has passed between preparation and measurement that the distribution of the walkers' movement has approximated their survival probability $\Psi(\tau)$.
We cannot necessarily expect the behaviour of these two cases to coincide as $\tau_0\to\infty$. This is because we are concerned with the time lag in the last run only, and not a sequence of runs which sum to this duration. There is thus a possibility of very long runs which do not necessarily reflect the equilibrium distribution. 

A considerable body of literature exists on time lags in random walk theory, be these Continuous Time Random Walks (CTRWs) \cite{agectrw, genctrw1, genctrw2, genctrw4, genrenctrw} or L\'{e}vy Walks specifically \cite{genlw1, genlw2}. Ageing in L\'{e}vy walks has been studied both in the context of MSDs, TAMSDs \cite{summaryreview, klafter} as well as the disparity between these two quantities and its relation to ergodicity breaking \cite{erg-lw1, erg-lw2}. Analogous results were previously found by Zumofen and Klafter in the context of dynamical systems \cite{zum1,zum2}, who also considered equilibrated and non-equilibrated initial conditions \cite{klafter}. The focus of this work is superdiffusive L\'{e}vy walks, though results may also be found in subdiffusive \cite{agesub} and ballistic \cite{polishage} regimes.
It has been suggested that ageing may lead to qualitatively different transport being observed \cite{lwage2, emergent}. Ageing may also play a role for L\'{e}vy walks in systems subject to finite time or space constraints as is indeed the case for empirical measurements of motility \cite{fin-time, fin-space}.
For an excellent review on time lags and their effects on random walks we refer the reader to \cite{summaryreview}, otherwise a comprehensive review specific to L\'{e}vy walks was published in \cite{levyreview}. A single equation for the L\'{e}vy walk has also recently been derived \cite{single}.
The structural probability density has previously been used to explore the effects of ageing on movement between discrete states \cite{prevage} or for jump processes \cite{genctrw2}. 
Our contribution is the joint study of ageing and stopping effects in the random walk, which yield different results than when studied separately. As will be shown, these results can have far-reaching implications for the interpretation of the underlying dynamics of the observed motion.

The structure of the paper is as follows: in the subsequent Section \ref{sec: framework} we develop the mathematical framework with which to describe ageing walkers, starting from the mesoscopic outset of $n(\vec{x},t,\tau)$. After summarising some key technical methodology, we derive the macroscopic equations for random walkers subject to fixed (see Section \ref{sec: const-lag}) and equilibrated (see Section \ref{sec: equil-lag}) time lags. In both cases we compare the results for the probability density of a single walker trajectory with equations for the mean density of walkers throughout space. The latter approach assumes a growing mean population of walkers and applies reaction-diffusion equations to predict the front velocity of the propagating group of walkers. We discuss the consequences for TAMSDs in Section \ref{sec: tamsd}. The results are discussed and concluded upon in Section \ref{sec: dandc}.

\section{General Description of Movement}
\label{sec: framework}

The aim of this section is to derive a general expression for movement of walkers via the structural probability density $n(\vec{x},t,\tau)$ under the influence of time lag. For simplicity, let us consider movement along the real line so the notation can be simplified to $\vec{v}=\pm v$ and $\vec{x}=x$.  We introduce the `turning rate' $\beta(\tau)=\frac{\psi(\tau)}{\Psi(\tau)}$ of the walkers which gives the rate at which a new velocity and running time is sampled. Notice that this rate may vary with $\tau$ to depict more or less persistent movement (decay or growth with $\tau$). If we separate the population of walkers according to their current direction of movement, we can write $n(x,t,\tau)=n_+(x,t,\tau)+n_-(x,t,\tau)+n_0(x,t,\tau)$, where $n_\pm$ denote the populations moving in the $\pm$ve directions (or populations with respective velocities $\pm v$). Similarly, $n_0(x,t,\tau)$ denotes the stopped walkers, though this quantity may be regarded as zero if only the motile walkers are of interest. The new component of the problem is that we will be taking into account that the walkers may have had runs which began before $t=0$, and that changes may occur in the total number of mobile walkers.

The walkers being measured are entities of finite lifetime (and possible reproduction), such that there is a replenishing rate $\eta$ of new walkers, and a stopping rate $\theta$. As already discussed, stops may result from a natural finite lifetime of the walker, premature death as a result of e.g. radiation damage during measurement \cite{tweezer2, tweezer1, tweezer3}, or binding to an alien component in the environment. An apparent stop of the walker need not correspond to `natural' death; we may simply be observing an experimental artefact or byproduct of the measurement. This is most apparent in the low photostability of photofluores employed as probes, leading to finite photon emission counts \cite{single3, photostable, revspt} and the detrimental effects of photobleaching \cite{bressloff}. Depending on the experiment in question, several of the suggested causes may be relevant concerns over multiple experiment iterations.
Circumventing these issues by the insertion of larger physical probes poses uncertainty regarding the perservation of walker behaviour, as well as the risk of accelerated sedimentation resulting from its increased size or mass \cite{revspt, sedimentation}. While concerted efforts have been made to reduce these issues and their effects on the sampled data \cite{fixbleach, sptfix, sptfix2}, they remain problematic and should not be ignored.\newline
The result is a net `stopping rate' which may include both technical noise, induced death from measurement and natural death of the walkers, though not all of these will be present in every experiment. Care must be taken when identifying this net rate as an effective death rate, as we thus overestimate the biological mortality of the walkers \cite{sptbd}. This problem becomes especially pertinent when studying walkers which merge, split and perish over the timespan of experimental observation. Having described the causes of walkers stopping, let us now proceed to their movement through space.

It is sensible to assume that there is no bias in the `newborn' or replenished walkers, and these thus begin moving in either direction with equal probability and zero initial running time. The equations of motion are thus given by
\begin{equation}
\frac{\partial n_\pm}{\partial t}\pm v\frac{\partial n_\pm}{\partial x}+\frac{\partial n_\pm}{\partial \tau}=-\beta(\tau)n_\pm-\theta n_\pm +\frac{\eta}{2}(n_++n_-).
\label{eq: start-meso}
\end{equation}
The unbiased direction of `newborns' is contained in the rate $n(x,t,\tau)/2$ rather than referring to a single direction $n_\pm$. In the case of a constant motile population with $\eta=\theta$, new walkers are replenished as often as others stop but need not move in the same direction as the walkers they are replacing. The overall movement of the walkers is consequently affected by the stopping of individuals. One can solve \eqref{eq: start-meso} using the method of characteristics and considering two cases: $\tau=t-t_s$ when $\tau<t$ and $\tau=t+\tau_s$ when $\tau>t$. The subscript $_s$ refers to the evaluation at the start of the characteristic. The general solution to the resulting equation $\int\frac{dn_\pm}{n_\pm}=-\int(\beta(u)+\theta)du$ is given by
\begin{equation}
n_\pm(x,t,\tau)=n_\pm(x_s,t_s,\tau_s)\exp\left(-\int_{\tau_s}^\tau\beta(u)+\theta du\right)
\end{equation}
where $\tau_s=0,\ \tau-t$ depending on whether $\tau<t$ or $\tau>t$, respectively. The values of $x_s$ and $t_s$ similarly depend on whether $t<\tau$ or $t>\tau$. Note the expression is independent of $\eta$ as replenished walkers appear with zero running time. The continued movement of random walkers is thus affected if a walker suddenly stops, cutting short the length of the current trajectory segment. Since the running time is reset for each segment, and the durations of these segments as dictated by $\beta(\tau)$ are independent of each other, the trajectories of replenished walkers are formally indistinguishable from the trajectories of `older' walkers after choosing a new direction of movement. 
Letting $n_\pm(x,0,\tau)=f^\pm(x,\tau)$, $n_\pm(x,t,0)=j_\pm(x,t)$, and choosing to write $\Psi(\tau)=e^{-\int_0^\tau\beta(u)du}=-\Psi'(\tau)/\beta(\tau)$, we obtain the solutions
\begin{equation}
n_\pm(x,t,\tau)=\begin{cases}
n_\pm(x\mp v\tau,t-\tau,0)\Psi(\tau)e^{-\theta\tau} & \text{if } \tau < t \\
f^\pm(x\mp vt,\tau-t)\frac{\Psi(\tau)e^{-\theta t}}{\Psi(\tau-t)} & \text{if } \tau > t.
\end{cases}
\label{eq: end-meso}
\end{equation}
We have identified $e^{-\int_0^\tau\beta(u)du}$ as the survival function $\Psi(\tau)$ corresponding to the turning rate $\beta(\tau)$. The result for the case $\tau<t$ is intuitively obvious: the probability density of the population at a certain point in time and space with given running time is the surviving portion of the population initially at position $x\mp v\tau$ and time $t-\tau$ with zero running time. As the walkers can also stop with a constant rate, only a certain portion $e^{-\theta\tau}$ remain after such a time interval. In the case of $\tau>t$ we must correct for the walkers which already started moving before this point, hence dividing by $\Psi(\tau-t)$.\newline
We are interested in the overall displacement of the walkers, for which it is convenient to introduce a macroscopic description $ P_\pm(x,t)$ of the walkers with velocity in either direction
\begin{equation}
 P_\pm(x,t)=\int_0^\infty n_\pm(x,t,\tau)d\tau,
\label{eq: defn-rho}
\end{equation}
which is the sum over all running times walkers may have whilst moving in the $\pm$ve direction. If no walkers stop, the above expressions sum to the PDF of a walker's position over time. As the walkers change direction, we consider a switching term $i_\pm(x,t)$ given by
\begin{equation}
i_\pm(x,t)=\int_0^\infty\beta(\tau)n_\pm(x,t,\tau)d\tau,
\label{eq: defn-i}
\end{equation}
which weighs these probabilities by the turning rates associated with each running time. $i_\pm$ is thus a probability flux corresponding to the change in direction of the walkers.
The initial conditions $f^\pm$ for the case $\tau>t$ are already given in \eqref{eq: time-conditions}, but we must provide the conditions for what happens when a run ends. The functions $j_\pm(x,t)=n_\pm(x,t,0)$ describe the walkers newly arriving at $x$ at time $t$ and starting a new run, and are given by
\begin{equation}
\begin{split}
j_\pm(x,t)=&\frac{1}{2}\int_0^\infty\beta(\tau)n_+(x,t,\tau)d\tau\\
&+\frac{1}{2}\int_0^\infty\beta(\tau)n_-(x,t,\tau)d\tau\\
=&\frac{1}{2}\left(i_+(x,t)+i_-(x,t)\right).
\end{split}
\end{equation}
By integration of \eqref{eq: start-meso} and using the above definitions we obtain the corresponding macroscopic evolution equation for $ P_\pm$ of the form
\begin{equation}
\frac{\partial P_\pm}{\partial t}\pm v\frac{\partial P_\pm}{\partial x}=\mp\frac{1}{2}[i_+(x,t)-i_-(x,t)]-\theta P_\pm+\frac{\eta}{2}( P_++ P_-).
\label{eq: macro-start}
\end{equation}
That is, the rate of change of walkers moving in either direction depends the probability flux of turning from either direction, as well as stopping-replenishing processes. 

In the context of e.g. cell division and death where $\eta,\ \theta$ may be interpreted literally, a stable population of walkers requires $\eta\geq\theta$. In such a case a probabilistic approach is less applicable, and a reaction-diffusion equation is considered to ascertain the velocity at which the mean population of walkers propagates. This is done in Sections \ref{sec: const-bulk} and \ref{sec: equil-bulk}.\newline 
However, the alternative interpretation where $\eta,\ \theta$ are e.g. stopping and replenishing rates of the random walkers lends itself to a probabilistic approach for a constant number of walkers. This requires that we either consider a further quantity $P_0(x,t)$, the immobile walkers across space (for a system where $\eta=0$), or let $\eta=\theta$ and renormalise for a constant number of motile walkers. For the stationary walkers, we find that
\begin{equation}
\frac{\partial P_0(x,t)}{\partial t}=\theta(P_+(x,t)+P_-(x,t)),
\label{eq: p0}
\end{equation}
such that $P_0$ increases locally where walkers cease to move. Assuming that initially all walkers are moving, it follows that $P_0(x,0)=0$. Note that \eqref{eq: macro-start} relies on quantifying the position and flux of the motile walkers. Using the definitions of \eqref{eq: defn-rho}-\eqref{eq: defn-i} combined with the integration of \eqref{eq: end-meso}, the motile walkers are described by
\begin{equation}
\begin{split}
 P_\pm(x,t)&=\int_0^tj_\pm(x\mp v\tau, t-\tau)\Psi(\tau)e^{-\theta\tau}d\tau\\
&+e^{-\theta t}\int_0^\infty f^\pm(x\mp vt,\tau)\frac{\Psi(\tau+t)}{\Psi(\tau)}d\tau
\end{split}
\label{eq: gen-rho}
\end{equation}
and (using $\psi=\beta\Psi$ from the definition of $\beta$) analogously the flux equals
\begin{equation}
\begin{split}
i_\pm(x,t)&=\int_0^tj_\pm(x\mp v\tau, t-\tau)\psi(\tau)e^{-\theta\tau}d\tau\\
&+e^{-\theta t}\int_0^\infty f^\pm(x\mp vt,\tau)\frac{\psi(\tau+t)}{\Psi(\tau)}d\tau.
\end{split}
\label{eq: gen-i}
\end{equation}
We notice that all information relevant to time lags is contained in the second term of the expressions for $ P_\pm,\ i_\pm$. Prehistory and future movement are thus separate in our analysis. These contributions decay exponentially due to the stopping of the walkers bringing forth these effects.

We must now ensure that the total probability of a walker occupying a certain position is commensurate with our different assumptions regarding the biological birth/death or stopping. If the walkers only change direction ($\eta,\ \theta=0$), the total probability is simply given by $P(x,t)=P_+(x,t)+P_-(x,t)$. If the number of walkers is preserved but some bind or cease to move with a rate $\theta$, then the total probability is instead
\begin{equation}
P(x,t)=P_+(x,t)+P_-(x,t)+P_0(x,t),
\label{eq: defn-tot-prob}
\end{equation}
where we additionally consider those walkers which over time attain zero velocity.\newline
In the interpretation where $\eta,\ \theta$ are viewed as birth and death rates the population of walkers is preserved when $\eta=\theta$. In such a context it makes little sense to consider the `dead' individuals as stationary walkers $P_0(x,t)$ since the moving walkers are replenished by the birth rate $\eta$. Instead the propagators for the moving, mortal walkers must be renormalised to accrue the effects of the `newborn' walkers which (on average) replace them.\newline
Naturally, for $\eta,\ \theta\ll1$ these rates will likely have little effect over short time scales, though this changes for longer times. The reasoning for this is as follows: $\theta$ results in shorter running times than those imposed should the walker have trajectories of durations specified by $\beta(\tau)$. The result is thus an increased `effective' rate at which a run ends, making longer excursions less likely. As $\theta$ grows, the duration of the last run of the walker will be prematurely shortened by the stopping or death of the walker. The effect of this changes whether we consider a single walker which stops ($\eta=0,\ \theta>0$), a renormalised probability of walkers which stop but are replenished ($\eta=\theta>0$), or a concentration of walkers where $\eta> \theta>0$.

If we consider a constant population, either in terms of binding or walkers which are replenished, we are concerned with a probability $P(x,t)$ of single trajectories as defined in \eqref{eq: defn-tot-prob}, and can be observed experimentally using single-particle tracking. 
However, subject to a trivial rescaling $P(x,t)$ becomes  a mean-field density $\rho(x,t)$ or concentration of walkers throughout space, which we call the `\textit{bulk}'. Here we are instead concerned with group motility of multiple walkers, where the total population on average is increasing. Single-particle tracking is not suitable for such an approach, but the group may be monitored by other means. As the walkers are non-interacting $\eta$ does not affect the probability of a single walker's trajectory, but will affect the bulk of all walkers (via the growth in the number of walkers).\newline
This distinction between single walker and bulk is important: for single-particle tracking we are concerned with the probability of a single walker's trajectory. However, study of the bulk may aid us in better interpreting the microscopically observed behaviour of each walker by the comparison of stopping rates of the bulk compared to individual walkers'. This may indicate whether $\theta$ is an inherent property of the walker (over the times that experiments are conducted) or an experimental artefact. Study of the bulk requires the execution of a separate experiment measuring the group propagation (specifically, the front velocity) as occurs when $\eta>\theta$ for sustained population growth. If this is not practicable, individual trajectories must suffice. The front velocity of a moving, living bulk of walkers has previously been studied in systems with no ageing \cite{SergeiBook, mmnp-1}. 
In this work we produce the corresponding bulk descriptions of walkers under ageing conditions, and outline the implications of these findings on the propagation velocity and interpretation of single trajectory observations. This shall be done for both initial conditions given in \eqref{eq: time-conditions}. In order to avoid confusion and to clarify the scope of validity of our results, calculations which specifically consider the bulk dynamics will be denoted using the mean-field notation $\rho(x,t)$ instead of $P(x,t)$. $P(x,t)$ thus refers to a random walker's PDF of position, and $\rho(x,t)$ the mean density of a growing population of random walkers that moves through space. Thus far our analysis makes no distinction between the two, as we must first construct the basic equations for the single walker before these can be averaged to describe the bulk. What follows is a brief description of the chosen methods by which to calculate our quantities of interest.

\subsection*{Brief Technical Interlude}
We are presented with somewhat involved integro-differential equations in \eqref{eq: macro-start}, which are often solved by working in Fourier or Laplace space. To this end, we introduce the Fourier-Laplace transform (FLT) of a function $g(x,t)$:
\begin{equation}
\mathcal{F}_x\mathcal{L}_t\{g(x,t)\}(k,s)=\tilde{g}(k,s)=\int_\mathbb{R}\int_0^\infty e^{ikx-st}g(x,t)dxdt.
\end{equation}
We also introduce the notation for a Fourier transform (FT) $\mathcal{F}_x\{g(x,t)\}(k,t)=\check{g}(k,t)=\int_\mathbb{R}e^{ikx}g(x,t)dx$ and a Laplace transform (LT) $\mathcal{L}_t\{g(x,t)\}(x,s)=\hat{g}(x,s)=\int_0^\infty e^{-st}g(x,t)dt$ of the function $g$. These transforms are very useful when working with integral definitions of functions, as illustrated here with the \textit{renewal measure} \cite{classicstats}.\newline
If we call each occasion a walker changes direction an `event', such that the mean number of events up to a time $t$ is given by $\mathbb{E}[N(t)]$, the renewal density is the rate of change of this quantity and one can write
\begin{equation}
h(t)=\frac{d\mathbb{E}[N(t)]}{dt}=\psi (t)+\int_{0}^{t}h(s)\psi (t-s)ds,
\label{eq: defn-h}
\end{equation}
more commonly known as the renewal equation for a running time PDF $\psi(\tau)$ of a walker \cite{renewal}. The benefit of these transformations is being able to trivially rearrange such integral equations to e.g. $\widehat{h}(s)=\frac{\widehat{\psi}(s)}{1-\widehat{\psi}(s)}=\frac{\widehat{\psi}(s)}{s\widehat{\Psi}(s)}$.

While powerful, this methodology is more easily applied by the separate consideration of the two forms of time lag $f^\pm$ posed in \eqref{eq: time-conditions}. In the following section we proceed the analyse the effects of each of these lags on the walker movement. 

\section{Some Examples of Time Lag}
In this section we obtain the propagators of single walkers and interpret these results by comparison with the bulk movement. This is done for both fixed and equilibrated time lags. To begin with, we shall consider the case of fixed running times $\tau_0$ at $t=0$, and determine the resulting MSD.

\subsection{Constant lag $f^\pm(x,\tau)=p_0(x)\delta(\tau-\tau_0)$}
\label{sec: const-lag}
Consider the effects of a time lag wherein the lag $\tau_0$ is fixed for all walkers. By substitution of $f^\pm(x,\tau)=p_0(x)\delta(\tau-\tau_0)$ into \eqref{eq: gen-rho}-\eqref{eq: gen-i}, we find the equations in Fourier Laplace space of the form
\begin{equation}
\begin{split}
\tilde{ P}_\pm(k,s)=&\tilde{\jmath}(k,s)\widehat{\Psi}(s\mp ikv+\theta)\\
&+\frac{\check{p}_0(k)}{\Psi(\tau_0)}\mathcal{L}\lbrace \Psi(t+\tau_0)\rbrace(s\mp ikv+\theta),
\end{split}
\label{eq: rho-const}
\end{equation}
and
\begin{equation}
\begin{split}
\tilde{\imath}_\pm(k,s)=&\tilde{\jmath}(k,s)\widehat{\psi}(s\mp ikv+\theta)\\
&+\frac{\check{p}_0(k)}{\Psi(\tau_0)}\mathcal{L}\lbrace \psi(t+\tau_0)\rbrace(s\mp ikv+\theta).
\end{split}
\end{equation}
These equations are similar, save for the presence of $\widehat{\psi}$ or $\widehat{\Psi}$. To relate these two quantities, we thus introduce the memory kernel $K(t)$ defined in Laplace-space as
\begin{equation}
\widehat{K}(s)=\frac{\widehat{\psi}(s)}{\widehat{\Psi}(s)}=\frac{s\widehat{\psi}(s)}{1-\widehat{\psi}(s)}=\frac{1}{\widehat{\Psi}(s)}-s=s\widehat{h}(s),
\label{eq: defn-k}
\end{equation}
where we have used the result that $\widehat{\psi}(s)=1-s\widehat{\Psi}(s)$. Eliminating for $\tilde{\jmath}(k,s)$ we can directly relate the probability flux to the probability of walkers where
\begin{equation}
\begin{split}
i_\pm(x,t)&=\int_0^tK(\tau)e^{-\theta\tau} P_\pm(x\mp v\tau,t-\tau)d\tau+\\
\frac{p_0(x\mp vt)e^{-\theta t}}{\Psi(\tau_0)}&\left\{\psi(t+\tau_0)-\int_0^tK(\tau)\Psi(t+\tau_0-\tau)d\tau\right\},
\end{split}
\label{eq: fixed-int}
\end{equation}
which is valid for all running time distributions $\psi(\tau)$. For $\tau_0=0$ the second term vanishes, as expected. The quantity $i_\pm(x,t)$ describes the flux of walkers moving in the $\pm$ve direction which change direction. It is the statement that the probability of newly arrived walkers equals the sum of probabilities of all walkers which were previously at positions with a velocity that now would allow them to be at position $x$ at time $t$.\newline
Noting that the above equations in Laplace space all have argument $s+\theta\mp ikv$, we introduce the shorthand $\omega_\mp=s+\theta\mp ikv$. From the walker flux \eqref{eq: fixed-int}, the definition of $j_\pm(x,t)=\frac{1}{2}[i_+(x,t)+i_-(x,t)]$ and \eqref{eq: rho-const}, it follows that
\begin{equation}
\begin{split}
2\tilde{ P}_\pm(k,s)&=\widehat{\psi}(\omega_\mp )\tilde{ P}_\pm(k,s)+\widehat{K}(\omega_\pm )\widehat{\Psi}(\omega_\mp )\tilde{ P}_\mp(k,s)\\
+\frac{\check{p}_0(k)}{\Psi(\tau_0)}\mathcal{L}&\lbrace \psi(t+\tau_0)*\Psi(t) - \psi(t)*\Psi(t+\tau_0) \rbrace(\omega_\mp )\\
+\frac{\check{p}_0(k)}{\Psi(\tau_0)}\mathcal{L}&\lbrace \psi(t+\tau_0)-K(t)*\Psi(t+\tau_0)\rbrace(\omega_\pm )\widehat{\Psi}(\omega_\mp )\\
+2\frac{\check{p}_0(k)}{\Psi(\tau_0)}\mathcal{L}&\lbrace \Psi(t+\tau_0)\rbrace(\omega_\mp ),
\end{split}
\label{eq: prop-const-ref}
\end{equation}
where $f(t)*g(t)=\int_0^tf(t-u)g(u)du$ denotes a convolution of two functions. A well-known approach in the literature \cite{klafter, pdf-start} is to use expressions for $\tilde{ P}_\pm$ to find the expression for the total probability $\tilde{ P}=\tilde{ P}_++\tilde{ P}_-+\tilde{P}_0$, where $\tilde{P}_0$ may be zero. From \eqref{eq: p0} we know that $s\tilde{P}_0(k,s)=\theta(\tilde{P}_++\tilde{P}_-)$ This is the propagator for the walkers already moving on the real line, which accounts for the rate $\theta$ at which the walkers stop. We thus find
\begin{equation}
\begin{split}
&\frac{\Psi(\tau_0)}{s+\theta}\frac{\tilde{P}(k,s)}{\check{p}_0(k)}=\frac{\mathcal{L}\lbrace \Psi(t+\tau_0)\rbrace(\omega_- )+\mathcal{L}\lbrace \Psi(t+\tau_0)\rbrace(\omega_+ )}{2s\check{p}_0|_{k=0}}\\
&+\frac{\widehat{\Psi}(\omega_+)\mathcal{L}\lbrace \psi(t+\tau_0)\rbrace(\omega_- )+\widehat{\Psi}(\omega_-)\mathcal{L}\lbrace \psi(t+\tau_0)\rbrace(\omega_+ )}{2s\check{p}_0|_{k=0}[2-\widehat{\psi}(\omega_+)-\widehat{\psi}(\omega_-)]}\\
&+\frac{\mathcal{L}\lbrace \Psi(t)*\psi(t+\tau_0)\rbrace(\omega_+)+\mathcal{L}\lbrace \Psi(t)*\psi(t+\tau_0)\rbrace(\omega_- )}{2s\check{p}_0|_{k=0}[2-\widehat{\psi}(\omega_+)-\widehat{\psi}(\omega_-)]},
\end{split}
\label{eq: prop-fixed}
\end{equation}
from which we can characterise the walker trajectories and MSD. This is the transformed PDF for the walkers moving through space, which is normalised for walkers which stop and are not replenished, or which are replenished at a rate $\eta=\theta$. 
Note that \eqref{eq: prop-fixed} does not apply to the total concentration of walkers, which is derived in a separate manner later in the text. 

A crucial point to note here is that usually one would not measure the MSD in a reaction-transport system containing processes such as death or binding (or general stopping of the motion). However, as already discussed, $\theta$ may affect the motion of the walkers by shortening longer trajectories, and must thus be considered. In other words, if we consider processes which eventually cease movement, there is a probability of shortened last segments of the trajectories due to an increased `effective rate' of changing direction. This becomes especially relevant if the tracked particles are in a constrained region or otherwise move in a system where disregarding the trajectory segments immediately before binding is unreasonable. Furthermore, for very persistent trajectories with occasionally long running times, disregarding the last segment in order to avoid issues pertaining to the binding may result in inaccurate measurements of the distribution of running times $\psi(\tau)$. Consequently, if the MSD is a quantity of interest which has been empirically measured, then a theoretical MSD as will be calculated below is of high relevance in predicting the variation from the MSD that would arise from similar dynamics in an `immortal' system (where practically $\theta=0$). If there are no expectations of constrained boundaries in the system such that binding is unlikely, experimentally measured MSDs of the form to be determined serve as indicators that internal cell dynamics or deterioration is at play in walkers of a size where directly observing such phenomena simultaneously is challenging.

Having motivated why an MSD for a stopping walker may be of interest, we now proceed to derive this quantity. We remind the reader that for such time-lagged PDFs, their Laplace transform is given by $\mathcal{L}\lbrace \psi(t+\tau_0)\rbrace(s)=\Psi(\tau_0)-s\mathcal{L}\lbrace \Psi(t+\tau_0)\rbrace(s)$, and use the well-known result that moments can be obtained from the characteristic function \cite{klafter, feller}
\begin{equation}
\mathcal{L}\lbrace\left<X^m(t)\right>\rbrace(s)=i^{-m}\left.\frac{\partial^m\tilde{ P}(k,s)}{\partial k^m}\right\rvert_{k=0}.
\label{eq: defn-var}
\end{equation}
As expected the mean position of the walkers $\left<X(t)\right>=0$ since the turning rates of the walkers are symmetric. The MSD thus requires us to evaluate:
\begin{equation}
\begin{split}
\left.\frac{\partial^2\tilde{ P}}{\partial k^2}\right\rvert_{k=0}&=-\frac{2v^2}{s}\frac{\mathcal{L}\lbrace \Psi(t+\tau_0)\rbrace(s+\theta)\widehat{\Psi}'(s+\theta)}{\Psi(\tau_0)\widehat{\Psi}(s+\theta)}\\
+\frac{2v^2}{s}&\left(\frac{\widehat{\Psi}'(s+\theta)}{(s+\theta)\widehat{\Psi}(s+\theta)}+\frac{\mathcal{L}'\lbrace \Psi(t+\tau_0)\rbrace(s+\theta)}{\Psi(\tau_0)}\right)\\
&+\frac{\check{p}''_0|_{k=0}}{s\check{p}_0|_{k=0}},
\end{split}
\label{eq: lapl-var-const}
\end{equation}
where the notation $'$ implies a derivative $\widehat{\Psi}'(s)=d\widehat{\Psi}(s)/d s$ in Laplace space and arises from the dependence of $\omega_\pm$ on $k$.
When stopping disappears ($\theta=0$), we recover the expected form arising from a L\'{e}vy walk with memory. Furthermore, in the case when the time lag is zero ($\tau_0=0$) we recover the known results for no ageing \cite{klafter, walkersrev}. However, once stopping comes into effect we find that the transport is tempered to produce a plateau once timescales become comparable to $\theta$. 
\newline
What occurs to this system for very large time lags, i.e. $\tau_0\to\infty$? We can let $\Psi(t+\tau_0)\approx\Psi(\tau_0)-t\psi(\tau_0)$ using $\psi(t)=-\Psi'(t)$. By manipulation of \eqref{eq: lapl-var-const} we can approximate
\begin{equation}
\begin{split}
\left.\frac{\partial^2\tilde{ P}}{\partial k^2}\right\rvert_{k=0}=&-\frac{2v^2}{s(s+\theta)^2}\left(1-\beta(\tau_0)\frac{\widehat{\Psi}'(s+\theta)}{\widehat{\Psi}(s+\theta)}-\frac{2\beta(\tau_0)}{s+\theta}\right)\\
&+\frac{\check{p}''_0|_{k=0}}{s\check{p}_0|_{k=0}},
\end{split}
\label{eq: beta-lag}
\end{equation}
where we have used the result that $\beta=\psi/\Psi$. For persistent random walks $\beta(\tau_0)\to0$ as $\tau_0\to\infty$. If this is the case, we find that for times $\tau_0\gg t$, the MSD of any walker is linear in time, but tempered by the stopping rate: $\left<X^2(t)\right>\sim te^{-\theta t}$. In the case when $\theta=0$, the MSD is ballistic with $\left<X^2(t)\right>\sim t^2$. However, other random walks may not be persistent, such that $\beta(\tau_0)\to\text{ const}$. In such cases, or for smaller time lags $t\geq \tau_0$, these terms in \eqref{eq: beta-lag} do not vanish and we must consider their effects. 

\subsubsection{Variances for Different Running Time Distributions}
We now evaluate the MSDs for different running times, subject to finite time lags $\tau_0<t$.

\textbf{Exponential Distribution:} $\psi(\tau)=\lambda e^{-\lambda \tau}$ which corresponds to a walker moving with a constant turning rate $\lambda$ and mean running time $\left<T\right>=1/\lambda$. We find that
\begin{equation}
\begin{split}
\left<X^2(t)\right>&=\frac{2v^2}{\theta\lambda(\theta+\lambda)} \left(\lambda+e^{-\theta t}[\theta e^{-\lambda t}-\theta-\lambda]\right)-\frac{\check{p}''_0|_{k=0}}{\check{p}_0|_{k=0}}\\
&\sim\quad t^0,
\end{split}
\label{eq: exp-const}
\end{equation}
which is constant over time. Considering that for non-stop transport ($\theta=0$) the MSD is linear in time, it should not be surprising that when $\theta>0$ the variance falls. The MSD is unchanged by time lag due to the constancy of the turning rate which indicates a Markov process. That stopping tempers the behaviour of the walkers is a well-known general effect previously seen in e.g. live cells \cite{sergeideath}. This is underscored by qualitatively different results arising from the walker mortality; an effect which can be seen when attempting to set $\theta=0$ in the above expression.

\textbf{Gamma Distribution:} $\psi(\tau)=\lambda^2\tau e^{-\lambda \tau}$ which has a mean running time of $\left<T\right>=2/\lambda$ and a preference for longer running times than the exponential distribution. In this case, for long times,
\begin{equation}
\begin{split}
&\left<X^2(t)\right>= \frac{2v^2}{1+\lambda\tau_0}t\frac{e^{-(\lambda+\theta)t}\lambda\tau_0}{\lambda+\theta}-\frac{\check{p}''_0|_{k=0}}{\check{p}_0|_{k=0}}\\
&+\frac{v^2 e^{-\theta t}}{\lambda(1+\lambda\tau_0)}\left(\frac{e^{-2\lambda t}(\lambda\tau_0-1)}{\theta+2\lambda}-\frac{3(1+\lambda\tau_0)}{\theta}\right),\\
&+\frac{v^2 e^{-\theta t}}{\lambda(1+\lambda\tau_0)}\frac{2e^{-\lambda t}[2(\lambda+\theta)+\lambda(2\lambda+\theta)\tau_0]}{(\lambda+\theta)^2},\\
&\sim\quad te^{-(\lambda+\theta) t}.
\end{split}
\label{eq: cont-single-gamma}
\end{equation}
A key feature of the transport here is that despite a nonzero stopping rate $\theta$ the MSD is linear in time, though tempered by the rates $\lambda,\ \theta$. Crucially, this is an ageing effect of walkers which vanishes when $\tau_0=0$. For sufficiently long times, the likelihood of such uninterrupted excursions decreases leading to a constant MSD analogous to an exponential distribution. In other words, ageing here results in a qualitatively different transport phenomenon than if one assumes movement commencing with zero running times.
Ageing or finite time effects have previously been found to result in variations in motility from subdiffusive to ballistic movement in systems with heavy-tailed running times \cite{fin-time} as we shall now investigate.

\textbf{Power Law Distribution:} $\psi(\tau)=\mu\tau_*^\mu/(\tau+\tau_*)^{1+\mu},\quad 1<\mu<2$ corresponding to a persistent random walk with a turning rate $\frac{\mu}{\tau+\tau_*}$ which decreases with the running time. This leads to a mean running time $\left<T\right>=\frac{\tau_*}{\mu-1}$ where $\tau_*>0$ is a characteristic time scale. 
Referring to \eqref{eq: lapl-var-const}, we need an expression for $\Psi(t+\tau_0)$. We recall that for no time shift and with characteristic time scale $\tau_*$, the survival function is given by
\begin{equation}
\Psi_{\tau_*}(t)=\left(\frac{\tau_*}{\tau_*+t}\right)^{\mu},
\end{equation}
where the subscript reiterates the time scale parameter. This notation will only be used where relevant for the rest of this section. The Laplace transform of the survival function is given by $\mathcal{L}_t\{\Psi_{\tau_*}(t)\}(s)=\widehat{\Psi}_{\tau_*}(s)=e^{s\tau_*}\frac{(s\tau_*)^{\mu}}{s}\Gamma(1-\mu,s\tau_*)$, where $\Gamma(\alpha,x)$ is the incomplete Gamma function \cite{rl-operator}. In the time lagged case we similarly find that
\begin{equation}
\begin{split}
\Psi_{\tau_*}(t+\tau_0)&=\left(\frac{\tau_*}{\tau_*+\tau_0}\right)^{\mu}\left(\frac{\tau_*+\tau_0}{\tau_*+t+\tau_0}\right)^{\mu}\\
&=\Psi_{\tau_*}(\tau_0)\Psi_{\tau_*+\tau_0}(t),
\end{split}
\label{eq: psi-lag}
\end{equation}
such that time lagged survival requires surviving throughout the lag $\tau_0$, as well as the subsequent time with a different time scale.
For ease of notation, let us introduce the constant $\gamma=\tau_*+\tau_0$. Then, $\mathcal{L}_t\{\Psi_{\tau_*}(t+\tau_0)\}(s)=\Psi_{\tau_*}(\tau_0)\widehat{\Psi}_{\gamma}(s)=\Psi_{\tau_*}(\tau_0) e^{s\gamma}\frac{(s\gamma)^{\mu}}{s}\Gamma(1-\mu,s\gamma)$. 
In the long-time limit (or equivalently when $s\to0$), we find that
\begin{equation}
\begin{split}
&\left<X^2(t)\right>=\frac{2v^2}{\tau_*^{1-\mu}}\frac{\Gamma(2-\mu)}{[\mu-1]^{-1}}\left(\frac{1}{\theta^{3-\mu}}-\frac{t^{3-\mu}Ei_{\mu-2}(\theta t)}{\Gamma(3-\mu)}\right)\\
&\qquad+\frac{2v^2}{\tau_*^{-\mu}}\left[\frac{\gamma^\mu}{\tau_*^\mu}-\frac{\gamma}{\tau_*}\right]\left(\frac{\Gamma(2-\mu)}{\theta^{2-\mu}}-t^{2-\mu}Ei_{\mu-2}(\theta t)\right)\\
&\qquad\sim\quad t^{3-\mu}Ei_{\mu-2}(\theta t),
\label{eq: res-fix-tail}
\end{split}
\end{equation}
where we have used the approximation $\widehat{\Psi}_{\tau_*}(s)\approx \left<T\right>+(s\tau_*)^{\mu}\Gamma(1-\mu)/s$ \cite{classicstats}, and the definition of the exponential integral $Ei_\alpha(x)=\int_1^\infty e^{-u x}/u^\alpha du$ \cite{rl-operator}. Hence, over shorter time scales the transport is superdiffusive with $\left<X^2(t)\right>\sim t^{3-\mu}$ which gradually becomes tempered to a constant MSD. This is consistent with the results obtained for previous running time distributions.
While the time lag does contribute over shorter time scales (of order $t^{2-\mu}$), the qualitative behaviour of the observed superdiffusion is unchanged by the time lag for larger times $t>\tau_0$. However, for systems where $\tau_0\gg t$ the MSD is tempered Brownian and follows $t e^{-\theta t}$. This is consistent with known findings from the literature \cite{genctrw4}.

\begin{figure}[h]
\includegraphics[width=0.47\textwidth]{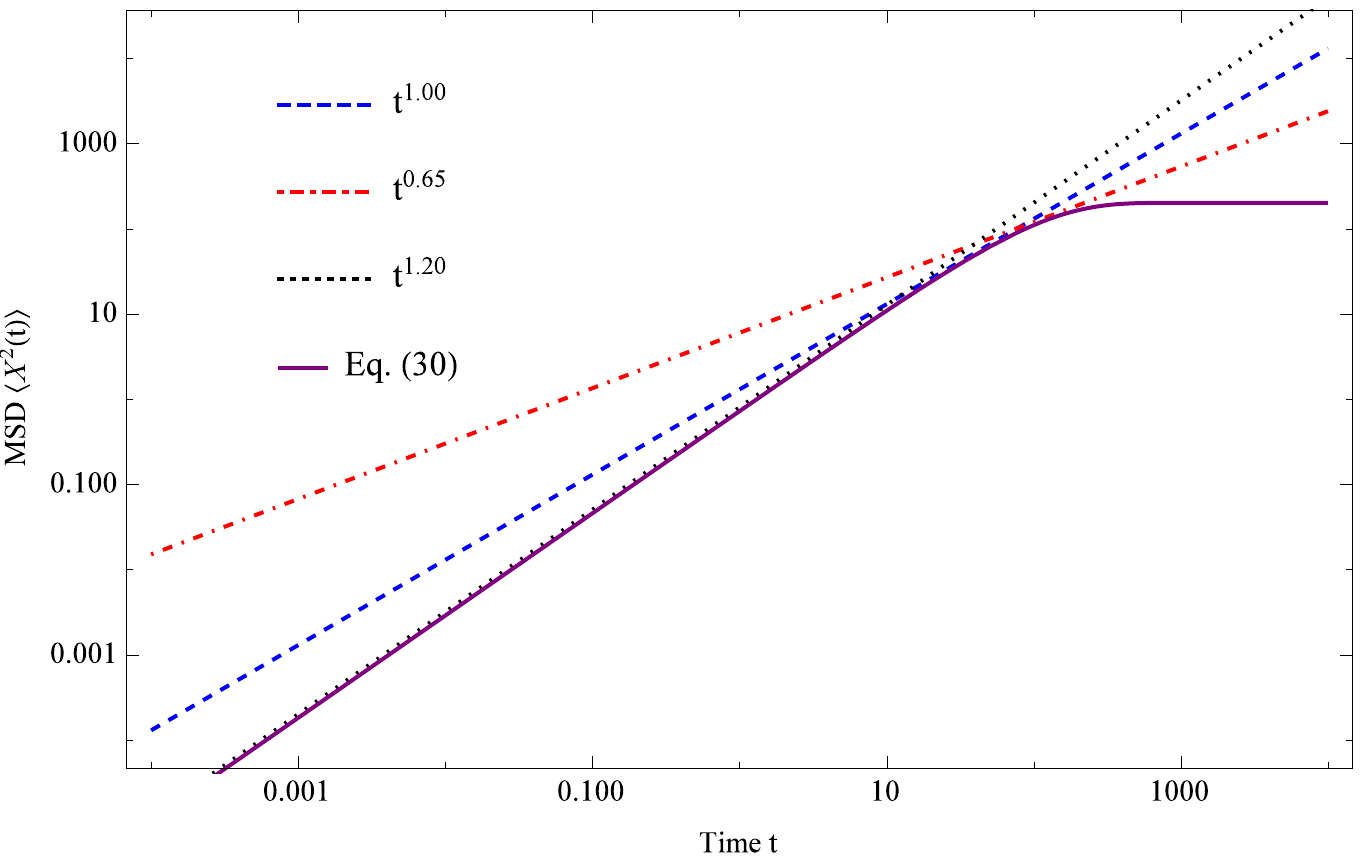}
\caption{Effect of tempering on an MSD with heavy-tailed running times. The bold line indicates a tempered superdiffusive process with $\mu=1.8,\ \theta=0.01,\ \tau_0=1$. Without tempering, this MSD would be expected to continue as indicated by the (black) dotted line. However, if measurements are taken over shorter time windows, there is a risk of identifying the resulting motion as Brownian (blue dashed line), subdiffusive (red dot-dashed line) or a joint process which undergoes all three stages. For clarity the subdiffusive component of \eqref{eq: res-fix-tail} has been suppressed.
\label{fig: eff-theta}}
\end{figure}

We have shown that the choice of running time PDF significantly influences the qualitative behaviour of the diffusion. In the gamma-distributed case the ageing is capable of producing a shift from one kind of diffusion to another. A shared feature is the tempering and eventual constant value of the MSD arising from the stopping rate $\theta$ as discussed in Section \ref{sec: framework}. 
We now briefly illustrate the effect of this tempering and resulting possible interpretations of the motility in Figure \ref{fig: eff-theta}.
If measurements are available over many orders of magnitude, one easily observes that there is a superdiffusive component which eventually reaches a plateau. Hence, the proposed model provides a natural mechanism for the presence of an `effective truncated power law' distribution of running times by the consideration of (natural or induced) death. Furthermore, for high-precision measurements which only record over shorter time intervals, there is a risk of interpreting the observed MSD as a transition from super- to subdiffusive motion as illustrated in Figure \ref{fig: eff-theta}. Care must thus be taken in arguing whether the movement simply appears to be behaving in a certain manner, or whether it is a reflection of the intrinsic movement of the walkers. Once the possibilities of experimental artefacts such as photobleaching or degeneration have been discarded, one can with greater confidence assert any transitions between movement regimes that might be observed.\newline
Naturally, when conducting the experiment a dataset will likely be discarded if little movement is seen (as corresponds to the expectations from a constant turning rate $\beta(\tau)=\lambda$). In the case of non-Markovian motion, measurement will likely also cease once it becomes clear that the walker is no longer moving (the plateau is reached). However, the times taken for this to occur will likely vary for each iteration of the experiment. 

Thus far, we have studied the MSD of single walkers moving throughout space as might be done using single-particle tracking. We now turn to a related system to be understood via a different experiment, with the intent that the results arising from these related investigations may bring greater understanding to walker motility. The walkers in question may still cease to move or die, but immobile walkers will be replaced with mobile ones at a rate $\eta\geq \theta$. If the number of walkers is not constant, it makes little sense to consider the MSD as a measure of the motility, and instead we study the mean-field density or bulk of walkers $\rho(x,t)$ throughout space. One can still think of the walkers as binding or dying as interchangeable interpretations, but unless $\eta=\theta$ there must now be a growth in the population of walkers in order to sustain the spread. 

\subsubsection{Description of the Bulk of Walkers}
\label{sec: const-bulk}

We shall now consider the mean population of walkers across space and time, which we term the bulk.
We must therefore include the birth of new walkers as these contribute to the population growth and hence the spreading of the bulk. The extent to which time lag contributes to this spread is also examined. Study of the bulk of walkers requires an experiment wherein a region of initially high concentration of walkers is monitored as the walkers multiply and spread into the empty surroundings, e.g. the movement of the boundary of a colony of walkers. The aim of such a description is to determine the front velocity $\vec{u}_\beta$ at which the bulk propagates into the unoccupied region, which may vary for each turning rate $\beta$. Heuristically, one further expects such a velocity to vary with the walker speed $v$, the birth rate $\eta$, and the death rate $\theta$, such that
\begin{equation}
|\vec{u}_\beta|=u_\beta(v,\eta,\theta).
\label{eq: u-parm-dep}
\end{equation}
The speed of the walkers is a constant for each system; the above expression simply underscores that it may be affected by the values of our chosen parameters as well as the motility of each walker. For each hypothesised turning rate $\beta$, the death rate $\theta$ may be estimated by inverting \eqref{eq: u-parm-dep} for known $\eta,\ v$. 
If one assumes the same parameter values for the bulk of walkers (e.g. $\theta_{\text{front}}$), and each individual random walker (e.g. $\theta_{\text{single}}$), comparison of these values may indicate whether further experimental artefacts arise in single-particle tracking should $\theta_{\text{single}}\neq \theta_{\text{front}}$. In other words, it can be used to determine whether the stopping rate of each walker is entirely determined by the death rate. We have implicitly assumed that the inherent movement (as given by $\beta$) undertaken by each member of the bulk is similar to what would be observed during single-particle tracking if there were no experimental artefacts. This approach thus only applies if walker interactions do not significantly alter motility.\newline
Having motivated the dual approach of single walker and bulk modelling, we now investigate the effects of ageing and mortality on the bulk propagation. Note that the employed methodology has previously been applied to estimate the speed of propagation during the Neolithic transition \cite{vlad02} and other non-Markovian processes (often with chemotactic interactions) \cite{h0, h02-2, mmnp-1}.

The starting point of this approach is still \eqref{eq: fixed-int}, but we now utilise the method popularised by Hillen and others \cite{h0, h02, h02-2, h0-2} to obtain a single integro-differential equation for the bulk (for derivation see Appendix \ref{sec: app-hillen}) where the population changes described in \eqref{eq: macro-start} are taken into account. 
By the algebraic manipulation of \eqref{eq: macro-start}, one obtains a single equation of the form:
\begin{equation}
\begin{split}
&\frac{\partial^2\rho}{\partial t^2}-v^2\frac{\partial^2\rho}{\partial x^2}+(2\theta-\eta)\frac{\partial\rho}{\partial t}-\theta(\eta-\theta)\rho=\\
&\frac{e^{-\theta t}}{\Psi(\tau_0)}v\frac{\partial}{\partial x}[p_0(x- vt)-p_0(x+ vt)]\times\\
&\quad\times\left( \psi(t+\tau_0) - \int_0^tK(\tau)\Psi(t+\tau_0-\tau)d\tau\right)\\
&-\int_0^t\frac{K(\tau)}{2}e^{-\theta\tau}\left(\frac{\partial}{\partial t}+\theta-\eta-v\frac{\partial}{\partial x}\right)\rho(x- v\tau,t-\tau)d\tau\\
&-\int_0^t\frac{K(\tau)}{2} e^{-\theta\tau}\left(\frac{\partial}{\partial t}+\theta-\eta+v\frac{\partial}{\partial x}\right)\rho(x+ v\tau,t-\tau)d\tau.
\end{split}
\label{eq: single-const}
\end{equation}
As before, when $\tau_0=0$ the ageing contribution vanishes and we have the standard transport equation for the bulk. A spatially uniform initial distribution $p_0(x)$ also leads to a vanishing ageing contribution. This is one means of establishing a basis of comparison between ageing effects and other (previously discussed) variables.
The macroscopic transport (from a mesoscopic outset) and its changes with ageing effects has previously been studied for walkers, though not taking into account their mortality \cite{agetransport}.

For groups of walkers it is of interest to determine the front velocity, i.e. the velocity at which the growing bulk of walkers spreads through space. In order to find the front velocity $u_\beta$, we apply a \textit{hyperbolic scaling} to the equation for the mean number of walkers \eqref{eq: single-const} which takes the form
\begin{equation}
t\to\frac{t}{\epsilon},\qquad x\to\frac{x}{\epsilon}
\end{equation}
and subsequently let $\epsilon\to0$ \cite{SergeiBook, fronttheory}. Crucially, the front velocity is unchaged by this as space and time coordinates are scaled equally. The contribution of ageing to this equation is enveloped by the death of these walkers $e^{-\theta t}$, such that when hyperbolic scaling is applied to these terms (leading to $e^{-\theta t/\epsilon}$) they vanish as $\epsilon\to0$. Further details may be found in \cite{mmnp-1} for similar random walks without ageing. The front velocity $u_\beta$ is upper bounded by the speed of walkers $u_\beta\leq v$, and decays to zero as the population stagnates $\theta\to\eta$.
Consequently, while the birth and death of walkers still needs to be taken into account for the calculation of the front velocity, it is a measure that is independent of ageing. Since the bulk motility as an ensemble average is less susceptible to errors and the birth of new walkers is unlikely to be an experimental artefact, the front velocity can be used as a gauge for the magnitude of these errors not inherent to the `life' of the walkers. Single trajectory expectations and the front velocity of the bulk in conjunction hence allow us to better gauge whether walkers are truly exhibiting e.g. Brownian motion, or only appear to be doing so due to the experimental effects previously discussed. 

We now proceed to investigate the effects of sufficiently long time lags that the system has equilibrated before measurement. We again consider both walkers which cease movement, and the movement of the bulk.

\subsection{Equilibrium state $f^\pm(x,\tau)=p_0(x)\frac{\Psi(\tau)}{\left<T\right>}$}
\label{sec: equil-lag}
We now consider the case wherein enough time has elapsed that the system has equilibrated. 
As more time passes, the running times thus become proportional to their survival probability $\Psi(\tau)$ scaled uniformly by the mean running time $\left<T\right>$ \cite{coxmiller, klafter}.
If the system starts from an asymptotic equilibrium state, we can write the time lag from \eqref{eq: gen-rho} as
\begin{equation}
\begin{split}
e^{-\theta t}\int_0^\infty& f^\pm(x\mp vt,\tau)\frac{\Psi(\tau+t)}{\Psi(\tau)}d\tau\\
&=\frac{p_0(x\mp vt)e^{-\theta t}}{\left<T\right>}\int_0^\infty\Psi(\tau+t)d\tau.
\end{split}
\end{equation}
Moving into Fourier Laplace space, equations \eqref{eq: gen-rho}-\eqref{eq: gen-i} become
\begin{equation}
\begin{split}
\tilde{ P}_\pm(k,s)&=\tilde{\jmath}(k,s)\widehat{\Psi}(s\mp ikv+\theta)\\
&+\frac{\check{p}_0(k)}{\left<T\right>(s\mp ikv+\theta)}(\left<T\right>-\widehat{\Psi}(s\mp ikv+\theta)),
\end{split}
\label{eq: rho-equil}
\end{equation}
and
\begin{equation}
\begin{split}
\tilde{\imath}_\pm(k,s)&=\tilde{\jmath}(k,s)\widehat{\psi}(s\mp ikv+\theta)\\
&+\frac{\check{p}_0(k)}{\left<T\right>(s\mp ikv+\theta)}(1-\widehat{\psi}(s\mp ikv+\theta)).
\end{split}
\end{equation}
By the same method as used in Section \ref{sec: const-lag}, we obtain
\begin{equation}
\begin{split}
i_\pm(x,t)&=\int_0^t K(\tau)e^{-\theta\tau} P_\pm(x\mp v\tau,t-\tau)d\tau\\
&+ p_0(x\mp vt)e^{-\theta t}\left\{\frac{1}{\left<T\right>}-\int_0^tK(\tau)d\tau\right\}
\end{split}
\label{eq: int-equil}
\end{equation}
as the expression for the walker flux. As done previously, we shall make use of the shorthand $\omega_\mp=s+\theta \mp ikv$ for arguments in Laplace space. From the walker flux \eqref{eq: int-equil}, the definition of $j_\pm(x,t)=\frac{1}{2}[i_+(x,t)+i_-(x,t)]$ and \eqref{eq: rho-equil}, it follows that
\begin{equation}
\begin{split}
2\tilde{ P}_\pm(k,s)&=\widehat{\psi}(\omega_\mp )\tilde{ P}_\pm(k,s)+\widehat{K}(\omega_\pm )\widehat{\Psi}(\omega_\mp )\tilde{ P}_\mp(k,s)\\
+\frac{\check{p}_0(k)}{\omega_\mp }&\left(2-\frac{\widehat{\Psi}(\omega_\mp )}{\left<T\right>}-\widehat{\psi}(\omega_\mp )\right)\\
+\frac{\check{p}_0(k)}{\omega_\pm }&\left(\frac{1}{\left<T\right>}- \widehat{K}(\omega_\pm )\right)\widehat{\Psi}(\omega_\mp ).
\end{split}
\end{equation}
In analogy to the statement regarding \eqref{eq: defn-tot-prob}, we may either consider a constant population of walkers where $\eta=\theta$, or interpret $\theta$ as a rate at which the walkers cease to move. 
The resulting probability of the walker moving through space, where the probability $P=P_0+P_-+P_+$ is then given by
\begin{equation}
\begin{split}
&\frac{(s+\theta)^2+k^2v^2}{s+\theta}\tilde{ P}(k,s)=\frac{\check{p}_0(k)}{\check{p}_0|_{k=0}}\frac{s+\theta}{s}+\\
&\frac{\check{p}_0(k)}{\check{p}_0|_{k=0}}\frac{ikv}{\left<T\right>}\frac{\widehat{\Psi}(\omega_+)(1-\widehat{\psi}(\omega_-))-\widehat{\Psi}(\omega_-)(1-\widehat{\psi}(\omega_+))}{s[2-\widehat{\psi}(\omega_+)-\widehat{\psi}(\omega_-)]},
\end{split}
\end{equation}
which can be used to characterise the motility of the walkers. We find from the symmetry of the turning rates that the mean displacement of the walkers is zero, but the MSD can be found from
\begin{equation}
\begin{split}
\left.\frac{\partial^2\tilde{ P}}{\partial k^2}\right\rvert_{k=0}=\frac{\check{p}''_0|_{k=0}}{s\check{p}_0|_{k=0}}-\frac{2v^2}{s(s+\theta)^2}\left(1-\frac{\widehat{\Psi}(s+\theta)}{\left<T\right>}\right).
\end{split}
\end{equation}
using \eqref{eq: defn-var}. We can now draw a comparison between this behaviour and that of the MSD for fixed time lags in \eqref{eq: lapl-var-const}. For random walks where the mean running time grows $\left<T\right>\to\infty$, we again encounter tempered Brownian behaviour with an MSD of order $t e^{-\theta t}$. However, when $\theta=0$ the MSD is ballistic; a phenomenon consistent with strongly anomalous motion ($0<\mu<1$), which also holds another interpretation. For random walks with large time lags $\tau_0\gg1$ as examined in Section \ref{sec: const-lag}, one empirically also measures very long running times if $\beta(\tau_0)\to0$. If the turning rate $\beta(\tau_0)\to\text{const.}$ we do not observe this behaviour, and thus have no reason to expect the motion to be ballistic. In the case of no stopping, very long fixed time lags can therefore be regarded as qualitatively equivalent to non-aged walkers which are strongly anomalous rather than superdiffusive. When $\theta>0$, the MSD of the walkers is slower and tapers off as the walkers bind. We now investigate the MSD for a variety of running time PDFs.

\subsubsection{Variances for Different Running Time Distributions}

\textbf{Exponential Distribution:} $\psi(\tau)=\lambda e^{-\lambda \tau}$ which corresponds to a walker moving with a constant turning rate $\lambda$ and mean running time $\left<T\right>=1/\lambda$. We find that
\begin{equation}
\begin{split}
\left<X^2(t)\right>&=\frac{2v^2}{\lambda\theta(\lambda+\theta)}\left(\lambda+e^{-\theta t}[\theta e^{-\lambda t}-\lambda-\theta]\right)-\frac{\check{p}''_0|_{k=0}}{\check{p}_0|_{k=0}}\\
&\sim\quad t^0,
\end{split}
\end{equation}
as expected. The MSD behaves the same as for no time lag due to the constancy of the turning rate which makes this random walk a Markov process. The result is therefore identical to that of \eqref{eq: exp-const}.

\textbf{Gamma Distribution:} $\psi(\tau)=\lambda^2\tau e^{-\lambda \tau}$ which has a mean running time of $\left<T\right>=2/\lambda$ and a preference for longer running times than the exponential distribution. In this case, 
\begin{equation}
\begin{split}
\left<X^2(t)\right>&=v^2\left(e^{-(\lambda+\theta)t}\left[\frac{t}{\lambda+\theta}+\frac{3\theta+4\lambda}{\lambda(\lambda+\theta)^2}\right]-\frac{3e^{-\theta t}}{\lambda\theta}\right)\\
&+\frac{3\lambda+2\theta}{\theta(\lambda+\theta)^2}-\frac{\check{p}''_0|_{k=0}}{\check{p}_0|_{k=0}}\\
&\sim\quad te^{-(\lambda+\theta) t}.
\end{split}
\label{eq: psi-single-gamma}
\end{equation}
This MSD also exhibits tempered Brownian behaviour before converging to a constant, which is consistent with our findings from \eqref{eq: cont-single-gamma}. Since ageing is an inherent assumption here from the statement of \eqref{eq: time-conditions}, there is no fixed time lag $\tau_0$ that may be set to zero to demonstrate that indeed the above behaviour arises from ageing. However, from inspection of \eqref{eq: cont-single-gamma} and \eqref{eq: psi-single-gamma} the qualitative agreement between the two should be taken as a clear indicator that ageing changes the expected observations in the system, regardless of the particular form in which it manifests itself.
In this case, the distribution of running times arising from an equilibrium condition at $t=0$ means that there is a high likelihood of trajectories which have longer excursions, and thus again we observe a time window in which stopping does not dominate the transport. 

\textbf{Power Law Distribution:} $\psi(\tau)=\mu\tau_*^\mu/(\tau+\tau_*)^{1+\mu},\quad 1<\mu<2$ corresponding to a persistent random walker with a turning rate $\frac{\mu}{\tau+\tau_*}$ which decreases with the running time. This leads to a mean running time of $\left<T\right>=\frac{\tau_*}{\mu-1}$ where $\tau_*>0$ is a characteristic time scale. In the long-time limit we find that the MSD is given by
\begin{equation}
\begin{split}
\left<X^2(t)\right>&=\frac{2v^2}{\tau_*^{1-\mu}}\Gamma(2-\mu)\left(\frac{1}{\theta^{3-\mu}}-\frac{t^{3-\mu}Ei_{\mu-2}(\theta t)}{\Gamma(3-\mu)}\right)\\
&\sim\quad t^{3-\mu}Ei_{\mu-2}(\theta t).
\end{split}
\end{equation}
This result is consistent with the literature \cite{genctrw4} in the case of walkers which never stop ($\eta,\theta=0$). The trajectories of single walkers appear less affected by initial conditions which commence from an equilibrated state than those which do not. This should not be surprising as equilibration assumes each trajectory to be representative of the underlying probability $\Psi(\tau)$ of the running times, and consequently little change is observed. For systems which commence with zero running times (the standard theoretical description for standard random walks) the MSD also does not coincide with the above result. 
We now proceed to investigate how equilibration of the system of walkers pre-measurement affects the bulk.

\subsubsection{Description of the Bulk of Walkers}
\label{sec: equil-bulk}
Using the same method as described in Appendix \ref{sec: app-hillen}, we obtain a single equation for the bulk probability of the form
\begin{equation}
\begin{split}
&\frac{\partial^2\rho}{\partial t^2}-v^2\frac{\partial^2\rho}{\partial x^2}+(2\theta-\eta)\frac{\partial\rho}{\partial t}-\theta(\eta-\theta)\rho=\\
&\left(\frac{1}{\left<T\right>}-h(t)\right)e^{-\theta t}v\frac{\partial}{\partial x}[p_0(x+ vt)-p_0(x- vt)]\\
-&\frac{1}{2}\int_0^tK(\tau)e^{-\theta\tau}\left(\frac{\partial}{\partial t}+\theta-\eta-v\frac{\partial}{\partial x}\right)\rho(x- v\tau,t-\tau)d\tau\\
-&\frac{1}{2}\int_0^tK(\tau)e^{-\theta\tau}\left(\frac{\partial}{\partial t}+\theta-\eta+v\frac{\partial}{\partial x}\right)\rho(x+ v\tau,t-\tau)d\tau.
\end{split}
\label{eq: total-intdiff}
\end{equation}
For initial spatial distributions $p_0(x)$ which are homogeneous the ageing effects vanish. This allows for the same basis of comparison as discussed for \eqref{eq: single-const}.
Note that time lag effects here are included by the renewal density $h(t)$ (defined in \eqref{eq: defn-h}), which is not unexpected. If the walkers are allowed to equilibrate before measurement, we must simply consider the rate of change of events relative to the mean.

For the same reasons as discussed after \eqref{eq: single-const}, the front velocity of the bulk in this case is also independent of the equilibrated ageing. We again find a front velocity $u_\beta\leq v$ which decays as $\theta\to\eta$. For further details on the front velocity in such systems, please see \cite{mmnp-1}.
A summary of the qualitative diffusion processes obtained in this work for ensemble-averaged MSDs can be found in Table \ref{tab: sum}.

\begin{table}[h]
\begin{center}
\begin{tabular}{ |c||c|c|c|  }
 \hline
 \multicolumn{4}{|c|}{Single walker $\left<X^2(t)\right>$ for different initial conditions} \\
 \hline\hline
 $\Psi(\tau)$& $\delta(\tau)$ &$\delta(\tau-\tau_0)$&$\Psi(\tau)/\left<T\right>$\\
 \hline
 $e^{-\lambda\tau}$   & $ t^0$   &$ t^0$&   $ t^0$\\
 $(1+\lambda\tau)e^{-\lambda\tau}$&   $ t^0$  & $ te^{-(\lambda+\theta) t}$  & $ te^{-(\lambda+\theta) t}$\\
 $\left(\frac{\tau_*}{\tau+\tau_*}\right)^\mu$ &$ t^{3-\mu}Ei_{\mu-2}(\theta t)$ & $ t^{3-\mu}Ei_{\mu-2}(\theta t)$&  $ t^{3-\mu}Ei_{\mu-2}(\theta t)$\\
 \hline
\end{tabular}
\end{center}
\caption{Qualitative description of the MSD for different running time PDFs and initial conditions. For large times all MSDs tend to a constant value, while the case $\theta=0$ recovers previously known results which are qualitatively different from the above. Global tempering of these quantities introduces the risk of na\"ive misinterpretation of the walker motility. Random walks with no memory (exponential) or strong memory (heavy-tailed) are either unchanged or slightly modified in the presence of ageing. Gamma-distributed MSDs are qualitatively different when we take into account ageing.}
\label{tab: sum}
\end{table}

Thus far we have considered the ensemble averages of trajectories which stop at a certain rate $\theta$ and, with exception of the gamma-distributed running time PDF, the results are qualitatively similar whether or not we take into account ageing. We now turn to the effect of stopping on another quantity of experimental interest, the TAMSD.

\section{The Effects on Time-Averaged MSDs}
\label{sec: tamsd}
One might argue that stopping only has perceptible effect on the ensemble average of trajectories. However, if all trajectory segments are of importance due to e.g. the dynamics of interest being close to the region where stopping occurs, we cannot disregard the segment during which stopping occurs. As a result, if the stop occurs at the end of the last segment, we need to consider how this may affect the conclusions drawn from our trajectories. For ease of illustrating the consequences of this effect, where necessary we shall disregard the effects of ageing in this last part of the trajectory. We expect the findings of this section to qualitatively coincide with equilibrated aged systems wherein a large portion of time has passed since preparation of the walkers and the experiment being carried out. Let us define the (ensemble averaged) TAMSD $\left<\bar{\delta^2}\right>$ to follow the definition
\begin{equation}
\left<\bar{\delta^2}\right>=\frac{1}{T-\Delta}\int_{0}^{T-\Delta}\left<[X(t+\Delta)-X(t)]^2\right>dt,
\label{eq: defn-tamsd}
\end{equation}
where the brackets $\left<\right>$ denote an ensemble average, $T$ is the total duration of measurement, and $\Delta$ the `windows' over which changes are observed \cite{klafter}. 
The TAMSD thus considers changes in the position averaged over the trajectory itself. The effects of binding on this are evident: if the trajectory contains few long segments, the `premature' shortening of one of these due to stopping can affect our conclusions. Naturally, if the trajectory consists of a very large number of segments, all of which describe motion in the region wherein the transport is of interest, it is reasonable to assume that one can discard the last trajectory segment. However, this may only be done if one is convinced the remaining set of trajectory segments are representative of the entirety of the motion being studied. Otherwise, the portion of the movement leading up to stopping must also be considered.\newline
In order to calculate $\left<\bar{\delta^2}\right>$, we require the correlation function for a L\'{e}vy walk, which can be calculated using the relation
\begin{equation}
\left<X(t)X(t+\Delta)\right>=\int_0^{t+\Delta}dy\int_0^{t}du\left<V(u)V(y)\right>
\label{eq: defn-pos-corr}
\end{equation}
where $V(u)$ is the velocity of the walker at a time $u$. Hence, the correlation of positions may be determined from the correlation of velocities, where we can make use of the fact that on the real line the velocity has the same magnitude but changes sign for every turn (every renewal event) the walker makes. Consequently, 
\begin{equation}
\left<V(u)V(y)\right>=v^2\sum_{n=0}^\infty (-1)^n p_n(u,y),
\label{eq: vel-corr}
\end{equation}
where $p_n(u,y)$ is the probability for $n$ turns in the time interval $(u,y)$. This problem was considered by Godr\`{e}che and Luck \cite{classicstats} and addressed for L\'{e}vy walkers without stopping in \cite{erg-lw1}. For a persistent random walker where $\beta(\tau)\to 0$ as $\tau\to\infty$, we can make the simplifying assumption that the most likely case is one wherein there are no turns, and can thus approximate \eqref{eq: vel-corr} to $\left<V(u)V(y)\right>\approx v^2p_0(u,y)$. In other words, this is the probability of uninterrupted movement (or persistence probability) during this time interval, during which the walker must also not stop.\newline
Let us consider the last segment of the trajectory, during which the movement stops. If $\theta=0$, continued movement would have occurred in a given direction for a duration which we denote $\tau_l$ (\textit{last} running time), sampled from the running time PDF $\psi(\tau)$. However, the final running time is instead $\tau_f\leq \tau_l$ which on average we expect to be less than $\tau_l$. The reason for this is simple: each previous segment of the trajectory were all at a risk of stopping with a probability $1-e^{-\theta t}$, such that over time the risk of stopping increases. Earlier movements occurred during a time where there was still a high probability of not stopping, and the durations of these segments were thus governed by the turning rates $\beta(\tau)$. However, the last segment of the trajectory has a larger probability of stopping, and the stop hence occurs at some point after this run has begun.

This argument can be illustrated with a heavy-tailed distribution of running times where the motion is superdiffusive and the survival probability $\Psi(\tau)=\tau_*^\mu/(\tau+\tau_*)^\mu$ is tempered by the stopping rate $\theta$. If we for simplicity consider the non-aged running times of the segment $\tau$ for motion which started at $t=0$, then the survival probability of the last segment is
\begin{equation}
\left(\frac{\tau_*}{\tau+\tau_*}\right)^\mu e^{-\theta t}.
\end{equation}
Each segment of the trajectory is independent of the next one as a new direction is chosen, but as time increases stopping of the walker becomes increasingly likely. Of course, if the total duration of measurement $T\ll t$ and $\theta\ll 1$, then the time scales over which stopping effects are likely to occur are beyond those of the experiment. However, for longer experiments or larger stopping rates, the mean duration of the final trajectory segment is
\begin{equation}
\begin{split}
\left<T_f\right>&=\frac{(\theta \tau_*)^\mu}{\theta}e^{-\theta (t_b-\tau_*)}\Gamma(1-\mu,\theta\tau_*)\\
&= e^{-\theta (t_b-\tau_*)}\left(\frac{\tau_*}{\mu-1}+\frac{(\theta \tau_*)^\mu}{\theta}\Gamma(1-\mu)\right)+\mathcal{O}(\theta\tau_*^2),
\end{split}
\end{equation}
where $t_b$ is the duration of time the walker was moving before the last segment of the trajectory began. For small $\theta\tau_*$, the mean duration $\left<T_f\right>$ contains the mean running time $\tau_*/(\mu-1)$ arising from heavy-tailed distributions with $1<\mu<2$. However, this mean time is diminished by the likelihood of the walker not stopping until now, which for larger $t_b$ may be significant.\newline
In order to determine the TAMSD of a trajectory which may be stopped at one end, \eqref{eq: defn-tamsd} requires that we calculate $\left<X^2(t)\right>$ and $\left<X^2(t+\Delta)\right>$ in addition to the correlation of the trajectory between two points. It is the last quantity defined in \eqref{eq: defn-pos-corr} where the presence of stopping is most notable as the shorter segment at the end of the trajectory is compared with positions at earlier times. There is thus a risk of the trajectories appearing less correlated than they actually are as $\Delta$ increases. The consequence of this is that, in analogy to the findings for the ensemble averaged MSD, the TAMSD slows in growth in a similar manner as illustrated in Figure \ref{fig: eff-theta}. 
The MSD and TAMSD for a superdiffusive L\'{e}vy walk have previously been related via a constant, indicating weak ergodicity breaking for such diffusion processes \cite{erg-lw1}. However, these two quantities are still qualitatively similar and this is expected to also be the case for the TAMSD. Specific details between the two experimental quantities may vary, but since stopping primarily affects displacement over longer times there is little reason to expect significant differences over shorter time scales. Over longer times, it is expected that the TAMSD will contain features such as slowing down in its growth (or even reaching a plateau) though the particular details will likely vary somewhat.

\section{Discussion and Conclusion}
\label{sec: dandc}

In this work the effects of ageing have been explored on random walks, with focus on superdiffusive subballistic movement which stops at a rate $\theta$, and thus has a finite timespan of movement. Ageing is applicable to any experiment involving single-particle tracking, and in most biological and physical systems it is understood that the duration movement will take place over is finite; the results are thus broadly applicable. The two forms of ageing (walkers all have a fixed non-zero initial running time, or have equilibrated to the survival function) have been shown to result in different diffusion processes for non-Markovian random walk processes. Certain subleading e.g. subdiffusive behaviour may arise in addition to diffusion coefficients which depend on the fixed time lag $\tau_0$.
All descriptions of the transport have been shown to slow down when considering a rate $\theta>0$, leading to a constant MSD of the walkers for sufficiently long times (heuristically consistent with time scales over which all walkers are expected to have ceased movement). In the case of gamma-distributed running times of the walkers, it has been shown that ageing produces a qualitatively faster spread of the walkers, resulting from the peak in the PDF $\psi(\tau)$. That is, ageing is shown to qualitatively counteract the slowing effects of binding over certain time windows, leading to an MSD $\left<X^2(t)\right>\sim t e^{-(\lambda+\theta) t}$. The observation of experimental quantities commensurate with this work's theoretical predictions thus provides a test by which to ascertain if binding, sedimentation, death or similar processes are occurring in systems where this is not directly observable.
Furthermore, we have shown that for sufficiently large time lags $\tau_0$, such that empirically the mean running time of the walker is very large (or theoretically $\beta(\tau_0)\to0$), the movement transitions from superdiffusive to ballistic in the case of $\theta=0$. We have further discussed the cases in which one may expect the MSDs from the fixed time lag and equilibrated initial conditions to qualitatively coincide. In the case where $\theta>0$, the MSD tends to $\left<X^2(t)\right>\sim t e^{-\theta t}$ when $\tau_0\gg t$; i.e. a variance consistent with Brownian motion tempered by the stopping rate.

In addition to ageing, we have considered the implications of the random walkers often having a finite duration in which they are motile before stopping. As a result, the diffusion processes are qualitatively altered due to the finite lifespan of the walkers, resulting in either a tempering factor $e^{-\theta t}$ or a plateau in the MSD once stopping/death dominates. The theoretical prediction of the transport close to times at which stopping occurs may be of interest in systems approaching such a state, as might occur in confined regions where binding may occur with the boundaries. 
As discussed, a portion of the apparent death effects may arise as experimental artefacts of observation, and in aid of this distinction we have included a mean-field description of the bulk of walkers which is expected to be less affected by external effects. The combination of the age-independent front velocity of the bulk with the predictions for individual trajectories should allow for improved understanding of the underlying dynamics of the walkers by the existence of a comparable measurement which is more robust in the face of experimental noise. The front propagation speed $u_\beta$ of the bulk should be easier to determine as it is less prone to the experimental issues discussed above and is directly related to the net growth rate $\eta-\theta$ of the walkers.

The implications of a stopping rate $\theta$ for a TAMSD has been discussed. As time (since the beginning of measurement) passes, there is an increasing likelihood of the walker stopping. As a result, when the walker finally stops the duration of this last trajectory segment is shorter than what one would expect from the running time PDF due to the `premature' stopping that $\theta$ causes. When calculating the TAMSD, it has been illustrated how such a shorter duration of the last trajectory segment may lead to a reduction in $\left<\bar{\delta}^2(\Delta)\right>$ for large $\Delta$. This is qualitatively similar to what was found for $\left<X^2(t)\right>$, and is of particular importance for running times which are heavy tailed and thus have a higher probability of long, uninterrupted trajectory segments.\newline
A key implication of this tempering is often neglected: namely the interpretation of the running times of the walkers in terms of optimal transport mechanisms when applied to e.g. intracellular transport. Experimental artefacts or constraints may result (for a na\"ive analysis) in seemingly exponentially distributed movement which is in fact described by e.g. a hidden underlying heavy-tailed distribution as illustrated in Figure \ref{fig: eff-theta}. This requires more careful analysis of future experimental results. In other words, we must consider whether the walker movement is inherently changing, or if it merely appears to be doing so due to a binding or stopping process.

We have made no attempt to separate the different causes of walker death, be these the result of intrinsic walker mortality or artefacts of the measurement. If the causes of mortality or stopping can be more precisely pinpointed via a function $\theta(t)$, it may be possible to separate these rates in the model and thus determine the times over which natural and experimentally imposed death rates dominate the observed motility. This could potentially identify the magnitude of systematic errors imposed by the measurements on the estimates of intrinsic walker behaviour and reproduction.

Future work aims at considering the effects of ageing on systems wherein the walkers are interacting, such that additional reactions may occur before observation begins. This may be of particular significance in the study of photosensitive reactions where the increased light intensity present under observation affects the reactions relative to their course during preparation of the sample.

\section{Acknowledgements}
The author wishes to thank S. Fedotov, T. Waigh and N. Korabel for helpful discussions on the topic.

\bibliographystyle{unsrt}
\bibliography{biblio}

\appendix
\numberwithin{equation}{section}
\section{Derivation of the General Equations of Time-Lagged Motion}
\label{sec: app-hillen}

The aim of this appendix is to outline the method, as popularised in \cite{h0, h02, h02-2, h0-2}, used to derive a single integro-differential equation for the bulk of walkers. In order to do so, we consider two quantities: $\rho(x,t)=\rho_++\rho_-$ and $J(x,t)=v(\rho_+-\rho_-)$. From \eqref{eq: macro-start} we thus find the expressions
\begin{equation}
\frac{\partial\rho_+}{\partial t}+ v\frac{\partial\rho_+}{\partial x}+\frac{\partial\rho_-}{\partial t}- v\frac{\partial\rho_-}{\partial x}=\frac{\partial\rho}{\partial t}+\frac{\partial J}{\partial x}=(\eta-\theta)\rho,
\label{eq: app-j}
\end{equation}
and 
\begin{equation}
\begin{split}
&\frac{\partial\rho_+}{\partial t}+ v\frac{\partial\rho_+}{\partial x}-\frac{\partial\rho_-}{\partial t}+ v\frac{\partial\rho_-}{\partial x}=\frac{1}{v}\frac{\partial J}{\partial t}+ v\frac{\partial\rho}{\partial x}=\\
&-[i_+(x,t)-i_-(x,t)]-\frac{\theta}{v}J.
\end{split}
\end{equation}
If we cross-differentiate the above expressions by $x, t$ we can eliminate $J(x,t)$ to find
\begin{equation}
\begin{split}
\frac{\partial^2\rho}{\partial t^2}-v^2\frac{\partial^2\rho}{\partial x^2}-\theta(\eta-\theta)\rho&+ (2\theta-\eta)\frac{\partial\rho}{\partial t}\\
&=v\frac{\partial}{\partial x} [i_+(x,t)-i_-(x,t)].
\end{split}
\label{eq: app-total}
\end{equation}
The right hand side of the above expression is most easily found for a particular set of initial conditions, and we shall consider both in what follows. The general expression can also be evaluated but is unnecessarily cumbersome for our intended use. It is helpful to note that $\rho_\pm=\frac{1}{2}\left(\rho\pm\frac{J}{v}\right)$. 
\subsection{Fixed Time Lag}
If the switching term for the walkers is given by \eqref{eq: fixed-int},
\begin{equation}
\begin{split}
&i_\pm=\frac{1}{2}\int_0^tK(\tau)e^{-\theta\tau}\left[\rho\pm\frac{J}{v}\right](x\mp v\tau,t-\tau)d\tau+\\
&\frac{p_0(x\mp vt)e^{-\theta t}}{\Psi(\tau_0)}\left(\psi(t+\tau_0)-\int_0^tK(\tau)\Psi(t+\tau_0-\tau)d\tau\right),
\end{split}
\end{equation}
then by differentiation (and using \eqref{eq: app-j}) we find
\begin{equation}
\begin{split}
&v\frac{\partial i_\pm}{\partial x}=v\frac{\partial p_0(x\mp vt)}{\partial x}\frac{e^{-\theta t}}{\Psi(\tau_0)}\psi(t+\tau_0)+\\
&\int_0^t\frac{K(\tau)}{2} e^{-\theta\tau}\left[v\frac{\partial \rho}{\partial x}\mp\frac{\partial\rho}{\partial t}\pm(\eta-\theta)\rho\right](x\mp v\tau,t-\tau)d\tau\\
&-v\frac{\partial p_0(x\mp vt)}{\partial x}\frac{e^{-\theta t}}{\Psi(\tau_0)}\int_0^tK(\tau)\Psi(t+\tau_0-\tau)d\tau.
\end{split}
\end{equation}
The explicit expression of $i_+-i_-$ substituted into \eqref{eq: app-total} yields \eqref{eq: single-const}.

\subsection{Equilibrated Time Lag}
On the other hand, if the switching term is given by \eqref{eq: int-equil}, such that
\begin{equation}
\begin{split}
i_\pm(x,t)&=\frac{1}{2}\int_0^t K(\tau)e^{-\theta\tau}\left[\rho\pm\frac{J}{v}\right](x\mp v\tau,t-\tau)d\tau\\
&+ p_0(x\mp vt)e^{-\theta t}\left(\frac{1}{\left<T\right>}-\int_0^tK(\tau)d\tau\right),
\end{split}
\end{equation}
then by an analogous method to the above we find
\begin{equation}
\begin{split}
&v\frac{\partial i_\pm}{\partial x}=v\frac{\partial p_0(x\mp vt)}{\partial x}e^{-\theta t}\left(\frac{1}{\left<T\right>}-\int_0^tK(\tau)d\tau\right)+\\
&\int_0^t\frac{K(\tau)}{2}e^{-\theta\tau}\left[v\frac{\partial \rho}{\partial x}\mp\frac{\partial\rho}{\partial t}\pm(\eta-\theta)\rho\right](x\mp v\tau,t-\tau)d\tau.
\end{split}
\end{equation}
This leads to \eqref{eq: total-intdiff}.

\end{document}